\documentclass{ws-procs9x6}

\usepackage{amssymb}
\usepackage{amsmath}
\usepackage{letterspace}
\usepackage{longtable}
\usepackage{textcomp}
\usepackage{array}

\begin{document}

\title{The SLAC Polarized Electron Source and Beam for \e158}

\author{T.~B. Humensky, for the \e158 Collaboration and the SLAC Polarized Electron Source Group}

\address{Dept. of Physics, UVa \\
E-mail: humensky@slac.stanford.edu}

\newcommand{\slsh}[1]{/\letterspace to -1.0\naturalwidth{#1}\ \ }
\newcommand{\bit}{\begin{itemize}}
\newcommand{\eit}{\end{itemize}}
\newcommand{\be}{\begin{enumerate}}
\newcommand{\ee}{\end{enumerate}}
\newcommand{\bd}{\begin{description}}
\newcommand{\ed}{\end{description}}
\def\alr{A_{LR}} 
\def\alrm{A_{LR}^{M\o ller}} 
\def\alrb{^{\text{beam}}\text{A}_{\text{LR}}} 
\def\alrbt{^{beam}\!\widetilde{A}_{LR}} 
\def\hcs{helicity correlations}

\def\z0{Z^0}

\def\e158{$\text{E-158}$}
\def\t437{$\text{T-437}$}
\def\hap{HAPPEX}
\def\lra{\leftrightarrow}
\def\mo{M\o ller}
\def\sw{\sin^2 \theta_W}
\def\q2{Q^2}

\def\NCA{{\em Nuovo Cimento} A}
\def\PHYS{{ Physica}}
\def\NPA{{ Nucl. Phys.} A}
\def\MATH{{ J. Math. Phys.}}
\def\PRO{{Prog. Theor. Phys.}}
\def\NPB{{ Nucl. Phys.} B}
\def\PLA{{ Phys. Lett.} A}
\def\PLB{{ Phys. Lett.} B}
\def\PLD{{ Phys. Lett.} D}
\def\PL{{ Phys. Lett.}}
\def\PRL{Phys. Rev. Lett.}
\def\PREV{ Phys. Rev.}
\def\PREP{ Phys. Rep.}
\def\PRA{{ Phys. Rev.} A}
\def\PRD{{ Phys. Rev.} D}
\def\PRC{{ Phys. Rev.} C}
\def\PRB{{ Phys. Rev.} B}
\def\ZPC{{ Z. Phys.} C}
\def\ZPA{{ Z. Phys.} A}
\def\ANNP{ Ann. Phys. (N.Y.)}
\def\RMP{{ Rev. Mod. Phys.}}
\def\CHEM{{ J. Chem. Phys.}}
\def\INT{{ Int. J. Mod. Phys.} E}
\def\NIMA{{Nucl. Instr. Meth. A}}






\thispagestyle{empty}
\renewcommand{\thefootnote}{\fnsymbol{footnote}}

\begin{flushright}
{\small
SLAC--PUB--9615\\
January 2003\\}
\end{flushright}

\vspace{.8cm}

\begin{center}
{\bf\large   
The SLAC Polarized Electron Source and Beam for \e158\footnote{Work supported by
Department of Energy contract  DE--AC03--76SF00515.}}

\vspace{1cm}

T. B. Humensky\\
Princeton University, Princeton, NJ 08544\\
for the \e158 Collaboration and the SLAC Polarized Electron Source Group\\
\end{center}

\vfill

\begin{center}
{\bf\large   
Abstract }
\end{center}

\begin{quote}
SLAC $\text{E-158}$ is making the first measurement of parity violation in M\o ller scattering.  $\text{E-158}$ measures the right-left cross-section asymmetry, $A_{LR}$, in the scattering of a 45-GeV polarized electron beam off unpolarized electrons in a liquid hydrogen target. $\text{E-158}$ plans to measure the expected Standard Model asymmetry of $\sim 10^{-7}$ to an accuracy of better than 10$^{-8}$.  This paper discusses the performance of the SLAC polarized electron source and beam during \e158's first physics run in April/May 2002.
\end{quote}

\vfill

\begin{center} 
{\it Contributed to} 
{\it International Workshop On From Parity Violation To Hadronic Structure And More ... } \\
{\it Mainz, Germany}\\
{\it June 5--June 8, 2002} 
\end{center}


\maketitle

\abstracts{
SLAC $\text{E-158}$ is making the first measurement of parity violation in M\o ller scattering.  $\text{E-158}$ measures the right-left cross-section asymmetry, $A_{LR}$, in the scattering of a 45-GeV polarized electron beam off unpolarized electrons in a liquid hydrogen target. $\text{E-158}$ plans to measure the expected Standard Model asymmetry of $\sim 10^{-7}$ to an accuracy of better than 10$^{-8}$.  This paper discusses the performance of the SLAC polarized electron source and beam during \e158's first physics run in April/May 2002.
}

\section{Introduction}
E-158 is making the first measurement of parity violation in M\o ller scattering by measuring the asymmetry in the cross section for scattering of longitudinally polarized electrons with an energy of 45 GeV off an unpolarized electron target:
\begin{equation}  A_{LR} = \frac{\sigma_R - \sigma_L}{\sigma_R +
\sigma_L},
\label{eq:asy}
 \end{equation}
where $\sigma_R$ ($\sigma_L$) is the cross section for incident right- (left-) helicity electrons.\cite{e158,cipanpbr}  \e158 will make a stringent test of the Standard Model and will consequently be sensitive to new physics.\cite{cmbrief} The expected Standard Model asymmetry is $\sim 0.1\ \text{ppm}$, and \e158 plans to measure it to better than $10\%$.  \e158 is described more completely elsewhere in these Proceedings;\cite{PaulPavi} here we focus on the polarized electron source, the beam performance, and the control of helicity-correlated asymmetries in beam properties, $\alrb$'s.\cite{humenskybr}

\section{Beam Delivery and Monitoring}
The SLAC accelerator division did an outstanding job delivering beam for \e158.\cite{turnerbr,deckerbr,woodleybr}  Table~\ref{tab:beam}  summarizes the beam properties achieved and the beam monitoring capabilities of \e158.  An important point to highlight is that the goals are driven by physics requirements.  Exceeding these goals leads to large gains in accelerator efficiency.  Two improvements in characteristics of the source laser system drove this accelerator performance.  First, the pulse-to-pulse intensity jitter of the laser beam was reduced from $1.5\%\ rms$ to $0.5\%\ rms$; because all  properties of the electron beam couple strongly to intensity, this resulted in greatly improved stability.  Second, there was a significant overhead of laser power available, allowing the temporal profile of the laser pulse to be shaped  to compensate for beam loading, reducing the energy spread within each pulse.

\e158 uses pairs of precision monitors of the beam intensity, energy, angle, and position  located in the beam line (A-Line) leading to End Station A.  A wire array just upstream of the target provides a measurement of the beam's intensity profile.  Additional monitors are located at the $1\text{-GeV}$ point in the accelerator.  The last four lines of Table~\ref{tab:beam} give the resolution of the beam monitors.  Our ability to monitor the beam intensity, energy, position, and angle as it approaches the target are close to or exceeding the requirements needed to achieve 1-ppb systematic errors on $A_{LR}$ due to uncertainties in these properties.
\begin{table}[th]
\tbl{Beam properties and monitor resolution for \e158 Run I.\vspace*{1pt}}
{\footnotesize
\begin{tabular}{|l|c|c|}
\hline
{} & Physics Goal & Run I Achieved \\
\hline
Electrons/pulse & $6 \cdot 10^{11}$ @ 45 GeV & $6 \cdot 10^{11}$ @ 45 GeV \\
 & $3.5 \cdot 10^{11}$ @ 48 GeV & $3.5 \cdot 10^{11}$ @ 48 GeV  \\
\hline
Pulse Length & $\sim 270$ ns & $\sim 270$ ns  \\
\hline
Repetition Rate & 120 Hz & 120 Hz \\
\hline
Charge Jitter & $2 \%\ rms$ & $0.5 \%\ rms$ \\
\hline
Position Jitter & 100 $\mu$m & 50 $\mu$m \\
\hline
Spot Size Jitter & $< 10\%$ of spot size & 5\% of spot size  \\
\hline
Energy Jitter & $0.2 \%\ rms$ & $0.03 \%\ rms$ \\
\hline
Energy Spread & $0.3 \%\ rms$ & $0.1 \%\ rms$ \\
\hline
Polarization & 75\% & (85 $\pm$ 5)\% \\
\hline
Efficiency & $\sim$ 50\% & $\sim$ (65-70)\%  \\
\hline
Position Mon. Resolution & 1 $\mu$m & 2 $\mu$m \\
\hline
Angle Mon. Resolution & 0.4 $\mu$rad & 0.1 $\mu$rad \\
\hline
Energy Mon. Resolution & 30 ppm & 40 ppm \\
\hline
Charge Mon. Resolution & 30 ppm & 60 ppm \\
\hline
\end{tabular}\label{tab:beam} }
\end{table}

\section{New Gradient-Doped Cathode}
The SLAC polarized electron source is based on photoemission from a GaAs cathode.\cite{Alley95br}  A new gradient-doped strained GaAsP cathode was installed prior to \e158's 2002 physics running.  This cathode (more fully described elsewhere\cite{takashibr}) was developed in a R\&D project for the Next Linear Collider (NLC) project.\cite{nlcbookbrief}  With the available laser power, this cathode can yield a charge of $2\cdot 10^{12}$ electrons in $100\ \text{ns}$ with no sign of surface charge limit.  This is significantly more charge than is required by \e158, providing additional flexibility in optimizing the optics system.  This cathode also provides polarization $> 80\%$.  Because this is a strained cathode, it has a  ``QE anisotropy,'' meaning that the quantum efficiency (QE) of the cathode depends on the orientation of the linear polarization of incident laser light.\cite{Mair96br} Typical ``analyzing powers'' are $5\text{-}15\%$.  This particular cathode has a relatively small analyzing power of $\frac{\Delta QE}{2\langle QE \rangle} \approx 3\%$.  The QE anisotropy is a dominant ingredient contributing to $\alrb$'s and its effects are discussed in section~\ref{sec:alrbs}.  

\section{Flash:Ti Laser System}
A flashlamp-pumped Ti:Sapphire laser drives the cathode's photoemission.  Table~\ref{tab:flashti} summarizes its parameters.\cite{brachmannbr}  The cavity output is an $\sim 15\ \mu\text{s}$ pulse.  A $270\text{-ns}$ pulse is sliced out of the region of lowest intensity jitter by a Pockels cell between crossed polarizers.  We shape the laser pulse's temporal profile using an additional Pockels cell--polarizer pair.  This pulse shaping is used to compensate for beam loading.  The new cathode provides peak electron polarization for an incident laser wavelength of $805\ \text{nm}$, as compared with $852\ \text{nm}$ for the previous standard GaAs cathode.  At $805\ \text{nm}$ the laser operates closer to the gain maximum of the Ti:Sapphire laser crystal, yielding a significant enhancement in laser performance.  Optimizing the choice of cavity end mirrors also improved performance.  These changes reduced the laser intensity jitter from $1.5\%\ rms$ to $0.5\%\ rms$.
\begin{table}[th]
\tbl{Parameters of the Flash:Ti laser beam as it ran for $\text{E-158}$ 2002 Physics Run I.  The position jitter at the photocathode is measured on the electron beam but is dominated by laser jitter.  The other entries are measured directly on the laser beam.\vspace*{1pt}}
{\footnotesize
\begin{tabular}{|l|l|}
\hline
Wavelength & 805 nm (Tunable over $750\text{-}850\ \text{nm.}$) \\
\hline
Bandwidth & $0.7\ \text{nm FWHM}$ \\
\hline
Pulse Length & 270 ns (Tunable over $50\text{-}500\ \text{ns.}$) \\
\hline
Pulse Energy & 60 $\mu$J (Typical.  Max. avail. is $600\ \mu\text{J}$ in $370\ \text{ns}$.)  \\
\hline
Circular Polarization & 99.8\%  \\
\hline
Energy Jitter & $0.5\%\ rms$  \\
\hline
Position Jitter at Photocathode & $<\ 70\ \mu\text{m}\ rms$ (For $1\ \text{cm}$ 1/e$^2$ diameter.) \\
\hline
\end{tabular}\label{tab:flashti} }
\end{table}

\section{Laser Polarization Control}
\label{sec:pita}
\begin{figure}
   \centering
   \begin{tabular}{c}
   \includegraphics[width=6cm]{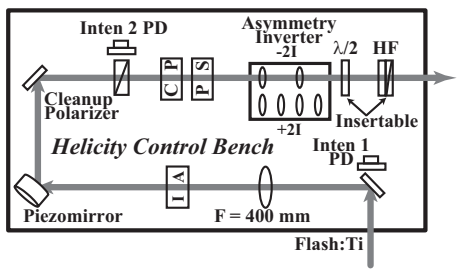}
   \end{tabular}
   \caption[example] 
   { \label{fig:hc} 
Optics for control of the laser beam's polarization and $\alrb$'s.}
   \end{figure}

Helicity correlations in electron beam parameters can in general be traced back to helicity correlations in laser beam parameters. The Helicity Control Bench optics (Fig.~\ref{fig:hc}) are designed to generate highly circularly polarized light of either helicity and to provide active means of controlling $\alrb$'s.

A linear polarizer (the ``Cleanup Polarizer") and a pair of Pockels cells (the ``Circular Polarization," or CP, and the ``Phase Shift," or PS, cells) control the polarization of the beam. The CP cell acts as a quarter-wave plate whose retardation flips sign pseudorandomly on a pulse-by-pulse basis, generating circularly polarized light of either helicity. The PS cell, pulsed at a low voltage, is used to compensate for residual birefringence in the optics between the Pockels cells and the cathode.  The phase shifts $\delta_{CP}$ and $\delta_{PS}$ induced by the CP and PS cells are proportional to the voltage applied to each cell; quarter-wave voltage is $\sim 2.6\ \text{kV}$.
The Cleanup Polarizer transmits horizontally polarized light.  The polarization at the exit of the PS cell can be expressed in the Jones notation as
\begin{math}
\overrightarrow{E} = \left(\begin{smallmatrix} \cos \delta_{CP}/2 \\ e^{i(\frac{\pi}{2}-\delta_{PS})} \sin \delta_{CP}/2 \end{smallmatrix}\right).
\end{math}
Appropriate choices of $\delta_{CP}$ and $\delta_{PS}$  allow the generation of arbitrary elliptically polarized light. We can parameterize imperfections in the laser polarization by writing $\delta_{CP}^{\pm} = \pm(\frac{\pi}{2} + \alpha_1)-\Delta_1,\ \delta_{PS}^{\pm} = \pm\alpha_2-\Delta_2$, where the superscripts $\pm$ indicate the two helicities.  It can be shown\cite{humenskybr} that the charge asymmetry depends linearly on the product of the QE anisotropy and the phases $\Delta_1$ and $\Delta_2$.  In particular, one can choose $\Delta_1$ and $\Delta_2$ such that the charge asymmetry is zero.  

\section{Sources of $\alrb$'s and Means of Suppression}
\label{sec:alrbs}
Several sources of $\alrb$'s are important for \e158.  These include
\be
\item Random fluctuations in the properties of the electron beam.  There are insufficient statistics in the experiment to average over random fluctuations until they are negligible at the level required by \e158.
\item Interaction between residual linear polarization in the laser beam ($\Delta_1$ and $\Delta_2$) and the cathode's QE anisotropy.  The average linear polarization across the face of the laser beam yields a charge asymmetry on the electron beam.  Spatial dependence of the polarization across the laser beam generates position and spot size asymmetries.
\item Helicity-correlated steering by the CP cell.
\ee

We use a number of strategies to control $\alrb$'s:
\be
\item The CP and PS Pockels cells give us complete control over the orientation and ellipticity of the polarization ellipse at the cathode, allowing us to achieve highly circularly polarized light.
\item The CP cell steering effect can be suppressed by imaging the CP cell onto the cathode.  Placing a beam waist at the CP and PS cells improves the uniformity of the laser beam's polarization.
\item Active feedbacks on the intensity and position asymmetries, utilizing the ``IA cell'' and the ``piezomirror'' in Fig.~\ref{fig:hc}, provide further suppression.  The feedbacks use beam monitors in the $1\text{-GeV}$ region of the accelerator in order to balance the properties of the beam between left and right helicities before accelerating to high energy.
\item Two methods of slow helicity reversal are used to flip the sign of the physics asymmetry and provide some cancellation of $\alrb$'s.  They are inserting a half-wave plate into the laser beam, which reverses the laser helicity; and running the electron beam at two energies, 45 and 48 GeV, in order to take advantage of the $g-2$ spin precession in the A-Line bend.  In addition, the ``Asymmetry Inverter'' shown in Fig.~\ref{fig:hc} flips the spatial profile of the laser beam leaving the CP and PS cells, and in principle provides some cancellation of position and angle asymmetries arising from them.
\ee

\section{Measurements of $\alrb$'s}
Table~\ref{tab:results} presents preliminary results on the control of $\alrb$'s during \e158 Run I.\cite{mastromarinobr}  These results are based on approximately 85 million pairs, representing roughly $25\%$ of the proposal data set.  The dependence of the detector asymmetry on $\alrb$'s was measured to be $\sim 1\ \text{ppb/ppb}$ for energy, $\sim 1\ \text{ppb/nm}$ for position, and $\sim 2\ \text{ppb/nm}$ for angle.  Typical corrections for each $\alrb$ are $\lesssim 10\ \text{ppb}$ and are known to $\sim 10\%$.  The corrections are below the level of the Run I statistical error ($\sim 25\ \text{ppb}$) and the uncertainty in $\alrb$'s contributes $< 10\ \text{ppb}$ to the Run I error.
\begin{table}[th]
\tbl{Preliminary results on $\alrb$'s for \e158 Run I.\vspace*{1pt}}
{\footnotesize
\begin{tabular}{|l|c|c|}
\hline
{} & Goal & Run I Achieved \\
\hline
$A_Q$ & 200 ppb & Target:  -280 $\pm$ 310 ppb \\
& & 1 GeV:  -96 $\pm$ 320 ppb \\
\hline
$\delta(A_Q)$ & 1 ppb & -2.4 $\pm$ 5.5 ppb  \\
\hline
$A_E$ & 20 ppb & 3 $\pm$ 13 ppb \\
\hline
$\delta(A_E)$ & 1 ppb & -2.0 $\pm$ 2.2 ppb \\
\hline
$(\Delta x, \Delta y)_{position}$ & 10 nm, 10 nm & -13 $\pm$ 4 nm, -3 $\pm$ 3 nm \\
\hline
$\delta(\Delta x, \Delta y)_{position}$ & 1 nm, 1 nm & 1.4 $\pm$ 0.6 nm, 1.6 $\pm$ 0.9 nm \\
\hline
$(\Delta x, \Delta y)_{angle}$ & 5 nm, 5 nm & 2 $\pm$ 5 nm, 1 $\pm$ 2 nm \\
\hline
$\delta(\Delta x, \Delta y)_{angle}$ & 0.5 nm, 0.5 nm & -2.1 $\pm$ 1 nm, 0.7 $\pm$ 0.6 nm \\
\hline
$(\Delta x, \Delta y)_{spotsize}$ & 10 nm, 10 nm & 1.4 $\pm$ 1.4 nm, 0.2 $\pm$ 1.5 nm \\
\hline
\end{tabular}\label{tab:results} }
\end{table}

\section{Conclusion}
\e158 has completed a successful first physics run during which the quality of beam delivered to the experiment was outstanding.  A new gradient-doped cathode provided both very high charge and polarization.  Improved laser performance led to $0.5\%$ charge jitter on the electron beam and very stable, efficient beam delivery.  Lastly, right-left beam asymmetries ($\alrb$'s) are small, yielding small corrections to the physics asymmetry.

\section*{Acknowledgements}
This work was supported in part by DOE contract DE-AC03-76SF00515.

\bibliography{p}
\bibliographystyle{unsrt.bst}
\end{document}